**Title**

Insula Interoception, Active Inference and Feeling Representation

**Authors**

Alan S. R. Fermin[1*], Shigeto Yamawaki[1], Karl Friston[2]

**Affiliation**

1 Center for Brain, Mind and Kansei Sciences Research, Hiroshima University, Hiroshima, Japan
2 The Wellcome Centre for Human Neuroimaging, UCL Queen Square Institute of Neurology, London, England.
*Correspondence: fermin@hiroshima-u.ac.jp

**Abstract**

The body sends interoceptive visceral information through deep brain structures to the cerebral cortex. The insula cortex, organized in hierarchical modules, is the major cortical region receiving interoceptive afferents and contains visceral topographic maps. Yet, the biological significance of the insula's modular architecture in relation to deep brain regions remains unsolved. In this opinion, we propose the Insula Hierarchical Modular Adaptive Interoception Control (IMAC) model to suggest that insula modules (granular, dysgranular and agranular subregions), forming networks with prefrontal (supplementary motor area, dorsolateral and ventromedial cortices) and striatum (posterior, dorsomedial and ventromedial) subregions, are specialized for higher-order interoceptive representations, recruited in a context-dependent manner to support habitual, model-based and exploratory adaptive behavior. We then discuss how insula interoceptive representations, or metaceptions, could give rise to conscious interoceptive feelings built up from low-order visceral representations and associated basic emotions located in deep interoceptive brain structures.

**Introduction**

The human brain is composed of various specialized brain regions for processing diverse functions such as problem-solving, decision-making, memory, motivation, inhibitory control, emotion, motor control, social, sensory, and somatosensory processing [1]. Recent studies have sought to understand the role of interoceptive visceral physiological processes under the control of the autonomic nervous system (ANS), in human behavior and disease [2-7]. The ANS is responsible for transmitting interoceptive information to the brain from visceral organs that serve to maintain body survival functions, including the gastrointestinal system, heart, lungs, liver, kidneys, thermoregulatory, hormonal, and immune systems [3, 4]. However, the functions for which cortical brain regions receiving interoceptive afferents are specialized and when such functions are recruited for interoceptive control remain unknown.

In the brain, the insular cortex is the main target of interoceptive afferents [3, 5, 8-11]. This central position of the insular cortex in processing interoceptive information has led to influential hypotheses on its potential roles in interoceptive prediction [9], information integration for awareness [12, 13] and interoceptive inference [14]. Despite the elegance and appeal of these hypotheses, no one has yet explained how insular functions are modulated by input from cortical and subcortical systems and whether such interactions maintain survival functions of the body. In this opinion we advance the Insula Hierarchical Modular Adaptive Interoception Control (IMAC) model to propose that the hierarchical organization of the insular cortex, supported by its multiple parallel networks with the prefrontal cortex (PFC), striatum, dopaminergic system, and autonomic nervous system, function in context-dependent control and learning of interoceptive responses, and in higher-order representation of conscious interoceptive feelings, built upon basic emotions and associated visceral processes.

This opinion starts with a brief review of basic neural pathways that connect the brain to visceral organs, and to the hormonal and immune systems, followed by a description of the modular cytoarchitectonic organization of the insular cortex and its parallel neuroanatomical networks with the cerebral cortex, striatum, and dopaminergic system. We then introduce the IMAC model and discuss how it explains interoceptive adaptive behavior and gives rise to higher-order representations, and possibly to conscious feelings. We conclude by identifying directions in which the concepts proposed by the IMAC model may be used to understand interoceptive dysfunctions observed in multiple psychiatric conditions, such as depression, addictive disorders, anxiety, post-traumatic stress disorder, and schizophrenia.

## Main Text

### Body-Brain Interoceptive Pathways

The pathways of the sympathetic nervous system (SNS) and parasympathetic nervous system (PNS) of the ANS convey afferent information about the physiological state of the body and visceral organs to the brain, and efferent central control signals to modify the activity of visceral systems [3, 4] (Figure 1). Interoceptive information also reaches the brain through the vascular system that transport all sorts of metabolites and signaling molecules, including nutrients, stress hormones, and immune substances, such as cytokines and prostaglandins [3, 15, 16].

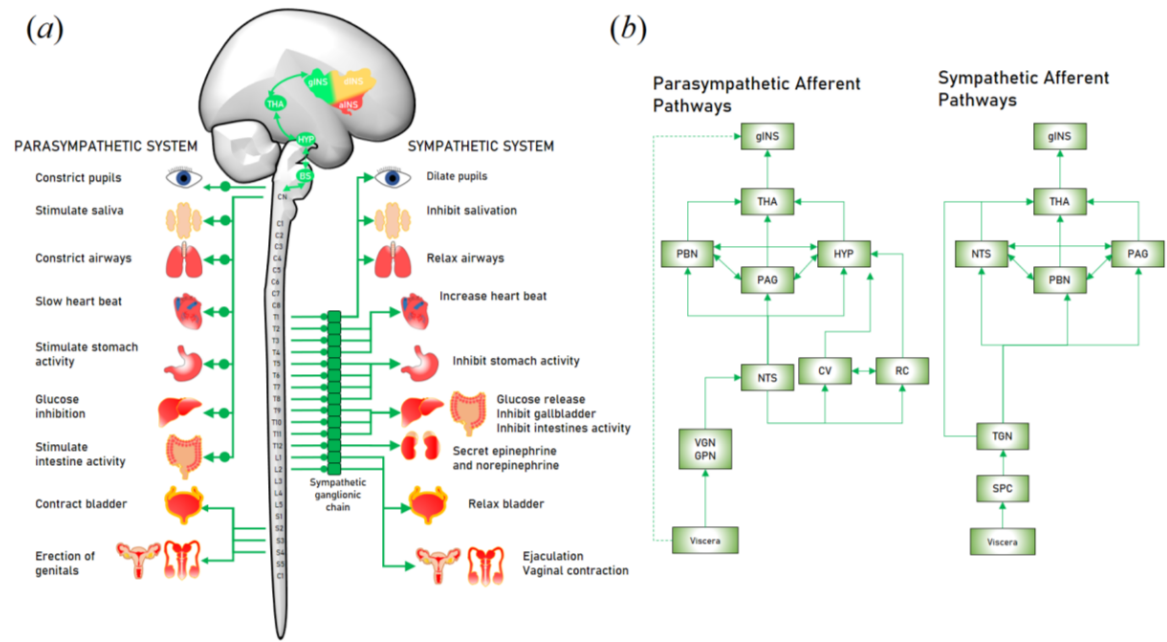

Figure 1. (*a*) Basic organization of the neural pathways linking the parasympathetic and sympathetic autonomic nervous systems to the central nervous system and the effects of their activations on visceral functions. (*b*) Schematic diagram of the parasympathetic and sympathetic afferent interoceptive neural pathways and their connections. In the parasympathetic scheme, the dashed line linking the viscera to the gINS represents circulatory (e.g. arteries, blood) interoceptive information. Abbreviations: gINS (granular insula), THA (thalamus), PBN (parabrachial nucleus), HYP (hypothalamus), PAG (periaqueductal gray), NTS (nucleus tractus solitarius), CV (cardiovascular reticular nuclei), RC (respiratory reticular center), VGN (vagus nerve), GPN (glossopharyngeal nerve), SPC (spinal cord).

Efferent pathways of both the SNS and PNS are organized in a two-neuron chain, the pre-ganglionic neurons that project to post-ganglionic neurons that synapse directly with their target visceral organs [3, 4]. Pre-ganglionic neurons receive input from the insular cortex and subcortical brain regions (e.g. amygdala, hypothalamus), and from brain stem systems including the reticular cardiovascular and respiratory nuclei [3, 4]. Cell bodies of PNS pre-ganglionic neurons are located in four main brain stem nuclei, the Edinger-Westphal nucleus (involved in pupil constriction), the salivatory nuclei (control lacrimal glands, salivary glands, and secretomotor glands of nasal and oral mucosa), the dorsal vagal nucleus (involved gut mobility and nutrient osmosis) and the nucleus ambiguus (innervates visceral organs and throat structures). PNS pre-ganglionic neurons project to post-ganglionic neurons via the cranial nerves, chiefly the glossopharyngeal (CN IX) and vagus (CN X) nerves [3, 4]. In contrast, the bodies of SNS pre-ganglionic neurons form viscerotopic maps located in the intermediolateral cell columns of the spinal cord, and project to their target post-ganglionic neurons located in the preaortic and sympathetic ganglia [3, 4].

Pre-ganglionic and post-ganglionic PNS neurons and pre-ganglionic SNS neurons release acetylcholine, whereas SNS post-ganglionic neurons that innervate smooth muscles release acetylcholine and SNS post-ganglionic neurons that innervate cardiac cells release noradrenaline[3, 4]. Acetylcholine binds to cholinergic and muscarinic receptors, and noradrenaline binds to four types of adrenergic receptors ($\alpha1$, $\alpha2$, $\beta1$, $\beta2$) [3, 4]. SNS and PNS post-ganglionic neurons also release distinct combinations of neuropeptides (e.g., somatostatin, substance P, dynorphin), forming a neurochemical code, specific to each target visceral organ, that modulates the speed and duration of visceral responses [3, 4, 17]. Activation of the SNS and PNS can exert antagonistic or synergistic influences, depending on the target visceral organs in question (See Figure 1A for examples of SNS and PNS influences on their visceral targets). An example of an antagonistic interaction is that SNS stimulation leads to pupil dilation and accelerates heart rate, but PNS stimulation leads to pupil constriction and decreases heart rate. In contrast, activation of both SNS and PNS act synergistically to increase secretion of lacrimal and salivary glands [3, 4].

Afferent ANS pathways transmit sensory information to the brain via nerve fibers connected to ANS efferent pathways [3, 4]. Most visceral information, such as glucose levels, blood oxygenation and osmolarity, heart contraction, and lung inflation and deflation, reaches the brain through the vagus nerve, 80% of which is composed of sensory fibers, whereas the SNS transmits information related to visceral pain [3, 4]. While cell bodies of SNS afferent neurons are located in the dorsal roots of the spinal cord, cell bodies of PNS afferent neurons are located in cranial sensory ganglia (Figure 1B) [3, 4]. After reaching the brain stem, SNS afferents synapse at the trigeminal nucleus (TGN) and from there to the nucleus tractus solitarius (NTS), parabrachial nucleus (PBN), and periaqueductal gray (PAG). Then from these systems, interoceptive sensory information is passed to the thalamus and finally reaches the insular cortex. Interoceptive afferent fibers travelling through the vagus nerve synapse at the NTS, and from there to the PBN, PAG, and hypothalamus, before arriving at the insular cortex, passing through the thalamus [3, 4].

**Insula hierarchical organization**

A series of neuroanatomical virus tracing studies in monkeys and humans have identified three main insula sub-regions based on the existence of a cellular layer IV containing predominantly granular cells: the agranular insula (aINS), located in a ventral anterior position and lacking layer IV; the granular insula (gINS), located in the most dorsal posterior position, with a fully developed layer IV, and the dysgranular insula (dINS), located between the aINS and gINS, but with an underdeveloped layer IV [18-20].

Insular sub-regions are organized hierarchically, with reciprocal connections between the gINS and dINS and between the dINS and aINS, and modest aINS fiber output to the gINS [18-20]. Another feature that characterizes the insular cortex is the progressive reduction of muscarinic acetylcholine receptors (rACh) along its ventral anterior to dorsal posterior axis, with a higher density of rACh in the aINS and a lower density in the gINS [18-20].

**Insula-visceral and somatic maps**

In monkeys, for example, vestibular, nociceptive, thermoreceptive, visceral, and gustatory information is represented in the insular cortex in a posterior to anterior direction [8]. Stimulation studies in humans have shown a topographic interoceptive representation of pain, thermal, somatosensory, visceral, and gustatory information located in a posterior to mid-insular direction [21]. In contrast, somatosensory information is represented all over the insular cortex [22, 23].

**Insula-cortical connections**

Cortical regions involved in primary sensory (somatosensory and auditory cortices), motor (primary, supplementary and pre-supplementary motor areas) and environmental information processing (superior and inferior parietal cortices) project predominantly to the gINS [8, 24]. It is important to note here that the inferior parietal cortex contains the supramarginal and angular gyri that form the temporo-parietal junction (TPJ), a cortical complex implicated in social cognition [25-27]. The dINS, on the other hand, receives anatomical input from the dorsolateral prefrontal cortex (Brodmann areas 45 and 46) [18-20, 28]. In contrast, the aINS makes connections predominantly with ACC, ventromedial orbitofrontal cortex, amygdala and hippocampal complex [8, 18-20]. Neuroimaging studies also show a topographic representation of emotional, motivational, cognitive, and viscero-motor information processing onto the insular cortex [29-34].

**Insula-striatum-dopaminergic connections**

The aINS makes connections primarily with the ventral striatum (vStriatum), the dINS with the dorsomedial striatum (mStriatum), and the gINS with the dorsolateral posterior striatum (pStriatum) [35]. Striatum sub-regions also receive topographical input from the dopaminergic system, with the vStriatum receiving its main dopaminergic input from the ventral tegmental area (VTA), and the mStriatum and pStriatum from the medial and ventro-lateral substantia nigra complex, respectively [36].

Striatum direct and indirect output pathways project to thalamic nuclei before reaching the insular cortex [37]. The medium spiny neurons (MSN) of striatum direct and indirect projection pathways express D1 or D2 dopamine receptors, respectively [38]. Both direct and indirect pathways project to the internal segment of the globus pallidus and substantia nigra pars compacta before reaching thalamic nuclei, however, the indirect pathway projects first to the external segment of the globus pallidus and the subthalamic nuclei [37, 38].

**Insula Hierarchical Modular Adaptive Interoception Control (IMAC) Model**

According to active inference, the brain uses internal generative models, acquired through experience, and extraction of the statistical distribution of sensory input, to continuously generate hypothetical predictions of expected sensory data [39-41]. In active inference, the goal of the agent is to find optimal policies that minimize free-energy, or prediction errors, between predictions of expected and actual sensory input generated by the agent's interactions with the environment [42, 43]. Based on active inference, the IMAC model hypothesizes that the hierarchical modular cytoarchitecture of the insular cortex, supported by its parallel neural networks with the PFC and striatum-dopaminergic systems, is specialized in higher-order interoceptive inference, hereby called metaception. That is, it specializes in formation of cortical interoceptive representations of lower-order innate interoceptive representations, hereby called mesaception, located in subcortical (e.g. amygdala, hypothalamus) and brain stem nuclei (e.g. reticular nuclei). In this framework, metaception is regarded as an evolved cortical capacity to generate higher-order interoceptive predictions that occur concurrently with action predictions that seek to maximize an individual's long-term fitness in his or her interactions with the environment. The main premise of the IMAC model is that adaptive behavior of animals and humans revolves around acquisition of behavioral policies, not only for stimulus-action mapping, but also for concurrent generation and learning of interoceptive responses associated with them.

Interoceptive sensory information first reaches the gINS, and is then forwarded to the dINS, and from the dINS to the aINS. In contrast, backward descending connections propagate information from the aINS through the dINS and back to the gINS. These forward and backward connections endow the insular cortex with a neuroanatomical architecture specialized for predictive coding, where backward connections generate and convey interoceptive predictions, and forward connections generate and convey interoceptive prediction errors [44, 45]. In this hierarchical insular architecture, the gINS represents the lowest level of the hierarchy, the dINS is at an intermediate level and the aINS sits atop the hierarchy. The gINS is in a position to generate low-order interoceptive predictions and to compute interoceptive prediction errors by comparing those predictions with real-time interoceptive afferents from the ANS. The dINS generates intermediate-order interoceptive predictions and computes interoceptive prediction errors based on forward signals arriving from the gINS. Finally, the aINS generates higher-order interoceptive predictions and interoceptive prediction errors by computing the difference between its predictions and the forward signals arriving from the dINS.

**Adaptive Interoception in Insula-Prefrontal-Striatum Networks**

The mechanism elaborated above is not entirely new, as previous works have already proposed how active inference may be used for interoceptive inference [9, 14]. The IMAC model, however, suggests that the insular interoceptive predictive functions can be better understood considering the parallel connections of the insular cortex with PFC sub-regions, namely the dorsolateral PFC (DLPFC), the ventromedial PFC (VMPFC), the supplementary motor area (SMA), and the striatum-dopaminergic system (Figure 2).

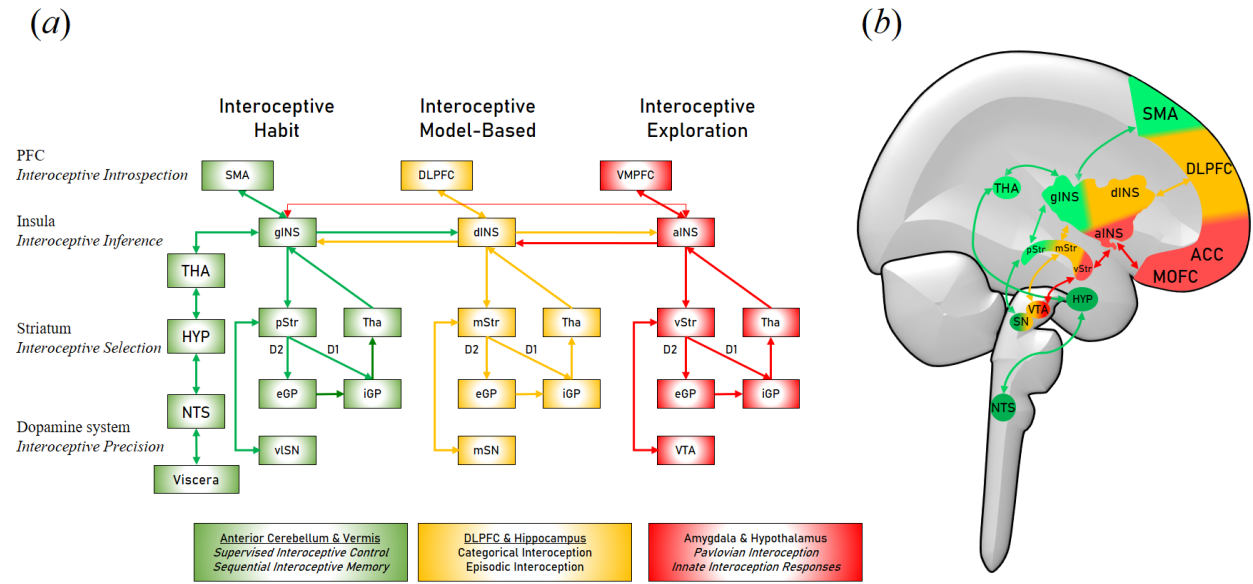

Figure 2. (*a*) Parallel neural networks linking the insula, prefrontal cortex and striatum. In the IMAC model, the PFC, insula, striatum and dopaminergic systems are specialized for distinct functions related to adaptive interoceptive inference, namely, interoceptive introspection, interoceptive inference, interoceptive selection and interoceptive precision, respectively. The insula-PFC-striatum parallel networks are hierarchically organized as follows: the lower level comprises the gINS-SMA-pStr network specialized for storage and quick generation habit-like interoceptive predictions; the mid-level comprises the dINS-DLPFC-mStr network specialized for model-based interoceptive predictions; and the highest level comprises the aINS-VMPFC-vStr network specialized for interoceptive exploration. Other neural, including the cerebellum, hippocampus and amygdala, provide additional functions to the interoceptive predictions of these networks, such as sequential interoceptive predictions, episodic interoception, and Pavlovian interoceptive conditioning, respectively. (*b*) Schematic representation of the neuroanatomical location of the main structures involved in adaptive interoceptive inference according to the IMAC model. Abbreviations: SMA (supplementary motor area), DLPFC (dorsolateral prefrontal cortex), VMPFC (ventromedial prefrontal cortex), ACC (anterior cingulate cortex), gINS (granular insula), dINS (dysgranular insula), aINS (agranular insula), pStr (posterior striatum), mStr (dorsomedial striatum), vStr (ventral striatum), eGP (external globus pallidus), iGP (internal globus pallidus), THA (thalamus), vlSN (ventrolateral substantia nigra complex), mSN (medial substantia nigra complex), VTA (ventral tegmental area), HYP (hypothalamus), NTS (nucleus tractus solitarius), D1 and D2 (dopamine receptors).

The PFC and striatum sub-regions form anatomical parallel loops specialized for processes implicated in adaptive decision making and behavioral control [37, 46-51]. The VMPFC-vStriatum and DLPFC-mStriatum loops are recruited in early stages of learning, when behavior is erratic, guided by external reward signals and require attention, but their activities diminish as learning progresses and behavior becomes automatic, fast, and less error-prone [46, 50-51]. Distinct neuroanatomical components of the anterior PFC and VMPFC-vStriatum loop are also implicated in novel learning by means of exploration [52-54], whereas the DLPFC-mStriatum loop implements a model-based, decision-making strategy to predict future outcomes of hypothetical actions [53, 55-57]. In contrast, activity of the SMA-pStriatum loop increases after repeated experiences, when behaviors become habitual. The SMA-pStriatum loop is implicated in sequential motor memory [46, 50, 51, 53].

The IMAC model hypothesizes that the metaceptive functions of the insular modules follows similar functional specializations as those observed in PFC-striatum loops. The posterior gINS, which forms a network with the SMA and pStriatum, is specialized for generating habitual interoceptive predictions using visceral-based representations. According to this view, the posterior gINS stores interoceptive memories, which can be readily used as interoceptive policies, especially in well-learned environmental situations, for fast generation of interoceptive predictions that control interoceptive responses. A selected gINS interoception policy is evaluated by dopaminergic precision signals arriving from the ventro-lateral SN complex and indicating the degree of confidence of the gINS in generating appropriate interoceptive predictions.

There are environmental demands, however, for which gINS interoceptive policies may be ineffective, eventually generating large interoceptive prediction errors. These gINS low-order interoceptive prediction errors are then passed forward and recruit metaception functions of the immediate insular module in the hierarchy, the dINS. The IMAC model hypothesizes that the dINS, as part of the DLPFC-dmStriatum loop, is specialized for model-based interoceptive predictions of interoceptive states that may be needed in future action states. Furthermore, dINS model-based interoceptive predictions may have a biological purpose to initiate allostatic interoceptive responses that adjust the functioning of visceral systems even before future action states or actual physiological disturbances take place [7]. A model-based strategy, however, has its limitations, given that it may actually be infeasible, due to high computational cost and time constraints, to predict all possible consequences of hypothetical actions and environmental state transitions. Conversely, a crucial advantage to the use of a model-based strategy is that it can significantly accelerate learning of novel behaviors [53, 56]. In the case of interoceptive behavior, hastening interoceptive learning in model-based adaptive behavior can be vital for acquisition and maintenance of interoceptive responses that are critical for survival.

Humans and animals constantly face novel situations that require learning of entirely novel behaviors, especially when old behaviors prove suboptimal and inefficient, or when internal models are outdated and unreliable. Recent evidence demonstrated that exploratory behavior, implemented in the VMPFC-vStriatum network, can be an alternative solution to support adaptive behavior in situations in which the SMA-pStriatum and DLPFC-dmStriatum fail to generate optimal behavioral policies [53, 56]. The IMAC model hypothesizes that the metaceptive function of the aINS, as part of the VMPFC-vStriatum network, is specialized for interoceptive predictions that support concurrent mapping of interoceptive responses to novel actions that proved successful in exploratory adaptive behavior. When model-based interoceptive predictions by the dINS fail, its interoceptive prediction error signals, modulated by dopaminergic precision signals in the medial SN complex, are sent forward to the aINS, which is then recruited to become dominant in exploratory, model-free behavior to support learning of novel interoceptive-state mappings

**Striatum, Dopamine, and Acetylcholine in Interoceptive Control**

Acetylcholine increases synaptic transmission in the thalamus, hippocampus, and PFC, making the activity of neurons in these regions more responsive to synaptic input from other brain areas, and facilitating learning of novel information [58-60]. The IMAC model hypothesizes that the higher density of acetylcholine receptors in the aINS and dINS endows these regions with greater capability to flexibly modify their metaceptive functions for interoceptive prediction and to create novel interoceptive mappings. In contrast, activity of the gINS can an exert direct influence on visceral interoceptive functions, such as generation of cardiac arrhythmias [61, 62]. The IMAC model hypothesizes that the lower density of acetylcholine receptors in the posterior gINS provides stability for interoceptive predictions that directly impact visceral survival functions.

The striatum has strong bidirectional connections with the dopaminergic system [36, 63]. The function of dopaminergic neurons is associated with value-based learning and signaling rewards, aversive cues, and alerting signals [64-66]. Striatal direct and indirect pathways have been implicated in facilitation vs inhibition of actions [67, 68], learning good vs bad values [69], and reward vs aversive learning, respectively [70]. Recent evidence, however, suggests that the direct and indirect pathways do not follow an on-off activation pattern, but rather a concurrent control of desirable and undesirable actions [71, 72]. In the context of active inference, the striatum is responsible for the selection of cortical representations based on precision signals computed by dopaminergic neurons [42]. The IMAC model hypothesizes that dopaminergic input onto the striatum serves to signal the precision or goodness of predictions arriving from the insular cortex, whereas the direct and indirect pathways are concurrently engaged in the selection of interoceptive responses that promote survival and suppression of responses that may cause harm.

**Stages of Insula Interoception Prediction Learning**

As described above, adaptive motor behavior goes through stages of learning and transitions of control in the brain [46, 50, 51, 53]. Similarly, the IMAC model hypothesizes that the aINS and dINS effect interoceptive predictions in early stages of learning that require exploratory or model-based interoceptive strategies, respectively, whereas the posterior gINS is recruited for interoceptive predictions once interoceptive behaviors have become habitual after repeated experiences. However, there are situations when learning by exploratory behavior is needed, but may eventually require learning of simple interoceptive responses. In such situations, it may be inefficient and life-threatening to use a more complex and time-consuming dINS model-based interoceptive strategy. The IMAC model hypothesizes that direct anatomical input from the aINS onto the gINS serves as an efficient neural pathway for fast learning of novel, but simple interoceptive predictions, such as in Pavlovian conditioning, which may support maintenance of visceral survival functions.

**Insula Feeling Representation**

In the IMAC framework elaborated above, two new terms have been proposed, metaception and mesaception, to explain how interoceptive predictions and interoceptive prediction errors give rise to neural representations of innate emotions and conscious feelings. Subcortical and brain stem systems store neural representations of unlearned motivational drives that are capable of generating innate behavioral repertoires (e.g., consummatory, approach, aggression) and automatic physiological reactions to interoceptive signals arriving from visceral organs [3, 4, 73-75]. Interestingly, innate behavioral repertoires are often associated with physiological responses and basic emotions, such as fear, anger, hunger, disgust, happiness, and surprise [75]. Thus, mesaceptions (or first-order representations) are interoceptive predictions computed in subcortical brain regions (e.g., amygdala, hypothalamus) and in brain stem nuclei (e.g., reticular nuclei, PAG, PBN) that give rise to basic emotions. For example, low glucose and insulin afferent signals onto brain stem systems activate a mesaceptive representation of hunger, and they trigger homeostatic responses and food-specific consumption behaviors through a specialized neural pathway within the hypothalamus [74, 76]. The hypothalamus has specialized survival systems involved in interoceptive processing that may give rise to other mesaceptions involved not only in emotions related to feeding, but also in emotions related to drinking, sex, aggression, fear, thermoregulation, and neural immunity [74].

Metaceptions or second-order interoceptive representations in the IMAC model, are higher-order cortical representations of mesaceptions, and they may generate conscious feelings brought about by visceral afferent signals and the basic emotions associated with them. Metaceptions may not only transform basic emotions such as fear and hunger into conscious feelings, but may also support the generalization of basic emotions into more complex emotional processes, such as social fear, embarrassment, pride, alertness, guilt, and grief. However, despite previous suggestions that that multiple cortical regions, including the insular cortex, contribute to the generation of conscious feelings [77-79], the specific role that these brain regions play in feelings remains elusive. The IMAC model offers a specific hypothesis that conscious feelings represented in the insular cortex are built from visceral signals and the innate emotions and behavior repertoires associated with them. Furthermore, based on the PFC-striatum networks in which each insula module participates, the IMAC model suggests that even within the insular cortex there may be different levels of feeling representation. The posterior gINS may generate implicit or habitual metaceptions commonly associated with "gut" or intuitive feelings, in contrast to the dINS and aINS that may contribute to representations of actual conscious or abstract interoceptive feelings.

The IMAC model suggests that metaceptions may initiate allostasis, that is, the anticipatory generation of interoceptive visceral responses to predicted future bodily and environmental demands[7]. However, a novel contribution of the IMAC model is the extension of the allostasis concept to suggest that metaceptions also have the capacity to generate feelings and modify interoceptive states when individuals replay past experiences, imagine hypothetical situations, empathize with others, and empathize with internal motivational states of others. For example, the insular cortex seems to possess mirror-neuron like functions that support empathetic behavior and understanding of others' feelings, which are then associated with physiological responses, such as crying when observing others crying, crying in grief, imagery of one's own and others' body sensations, and yawn contagiousness [80-84]. Furthermore, insula function is associated with music-induced feelings, art aesthetic judgment, interoceptive imagery, and retrieval of highly arousal, aversive, or disgusting experiences that have induced physiological changes [85-88].

The hierarchical organization of interoceptive feeling representation mapped onto the modular architecture of the insular cortex resembles the neuroanatomical and functional hierarchical organization of cognitive processes mapped onto the PFC and striatum [89, 90]. Functions of the PFC have also been associated with other forms of higher-order cognition, such introspection and reasoning [91]. Furthermore, stimulation studies in humans and animals have revealed that the SMA, DLPFC and VMPFC can induce physiological responses [92-94]. Based on the latter findings, the IMAC model hypothesizes that the PFC forms a third-order interoception representation and uses introspection to inspect and elaborate on the contents, causes, and consequences of second-order interoceptive representations of the insular cortex.

**Other brain interoception systems**
Several other brain regions may support the use of insular interoceptive representations to other domains (Figure 2). The DLPFC, which is also implicated in categorical learning [95, 96], may contribute to interoceptive introspection using categorical interoceptive information, such as heart rate, hunger level, or visceral pain. The hippocampus, implicated in episodic memory [97], may contribute to formation of episodic interoceptive memories and also to model-based interoceptive control [98]. The amygdala, implicated in Pavlovian conditioning [75], is connected with the aINS and may contribute to Pavlovian interoceptive learning. The neural architecture of the cerebellum has been implicated in computing forward models and encapsulation of learned sensory-motor mappings [48]. Distinct components of the cerebellar nuclei and vermis that function in visceral control and form loop connections with the SMA [99, 100], could store interoceptive forward models of sequential visceral responses.

(*a*) Common View

(*b*) IMAC Model

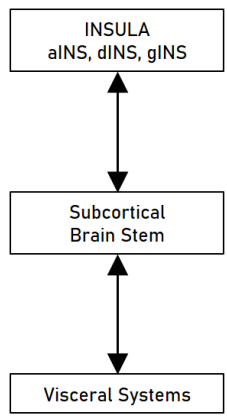
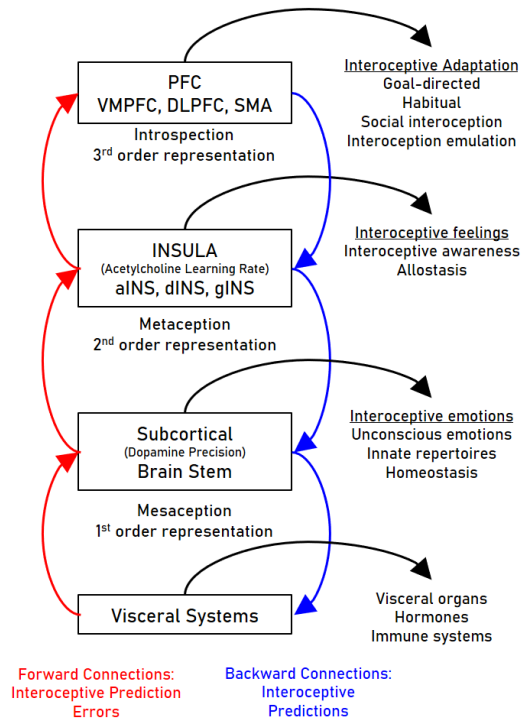

Figure 3. (*a*) Common view of interoceptive information flow ascending from visceral systems to subcortical regions and insular cortex and back to visceral systems. In this view, subcortical systems work as pattern generators and emit control signals to achieve homeostatic balance of set points. (*b*) The IMAC model proposes a hierarchical architecture where forward connections (red arrows) transmit interoceptive prediction errors and backward connections (blue arrows) generate interoceptive predictions of desired interoceptive states. In this hierarchical interoceptive architecture, deep brain systems (subcortical and brain stem areas) contain mesaceptions or 1st order representations of interoceptive information; the insular cortex contains metaceptions or 2nd order representations of mesaceptions; and the PFC contains 3rd order representations of interoceptive signals and implements introspection to guide adaptive interoceptive behaviors and estimate the meaning, causes and consequences of interoceptive prediction errors. (see Main Text for more details).

**Concluding Remarks**

In this opinion, we applied the active inference framework to propose an Insula Hierarchical Modular Adaptive Interoception Control (IMAC) model in an attempt to explain how the hierarchical modular organization of the insular cortex helps to form higher-order cortical interoceptive representations, adaptive interoceptive control, and interoceptive feelings. The IMAC framework contrasts with a view that takes for granted that unaltered interoceptive afferents reaching the central nervous system. The main contributions of the IMAC model can be summarized as four main points, as follows. Parallel networks linking the insular cortex with the PFC and striatum are specialized for hierarchical generation of interoceptive policies that map interoceptive predictions to particular behaviors in an experience-dependent manner. Second, the IMAC model suggests that dopamine emits precision signals regarding the goodness of insular interoceptive predictions and supports interoceptive predictions that support life and suppresses those that are harmful to the maintenance of visceral functions. Third, acetylcholine is hypothesized to support flexibility in formation of novel interoceptive mappings by the dINS and aINS, and maintains stability in the gINS of interoceptive predictions implicated in maintenance of visceral survival functions. Last, the IMAC model introduces two novel concepts, metaception and mesaception, to dissociate functions of interoceptive representations in the insular and subcortical / brain stem systems (Figure 3).

Altogether, the IMAC model takes a network approach to explain how higher-order interoceptive predictions, computed in the hierarchical modular organization of the insular cortex, give rise to conscious feelings built up from basic innate emotions represented in subcortical and brain stem systems. Future work is needed to test the predictions made by the IMAC model, as well as to test how disturbances in different components of the model, including metaceptions and mesaceptions, contribute to social anxiety and stress and lead to development of somatic and interoceptive dysfunctions observed in multiple neurological disorders, such as depression, or addictive or post-traumatic stress disorders.

**Acknowledgments**

A. F and S. Y. are grateful for research support provided to S. Y. by the Japan Agency for Medical Research and Development (AMED), Grant No. JP19dm0107093, and the COI

STREAM program from the Japan Science and Technology Agency (JST), Grant No. JPMJCE1311.

**Competing Interests**

The authors declare no competing interests.

**References**

1.  Kandel, E. R., Schwartz, J. H., & Jessell, T. M. Principles of neural science. New York: McGraw-Hill, Health Professions Division (2000).

2.  Porges, S. W. The polyvagal theory: Neurophysiological foundations of emotions, attachment, communication, and self-regulation. New York etc: Norton (2011).

3.  Jänig, W. The integrative action of the autonomic nervous system: Neurobiology of homeostasis. Cambridge: Cambridge University Press (2008).

4.  Wilson, P. L., Stewart, P. A., & Akesson, E. J. Autonomic nerves: Basic science, clinical aspects, case studies. Philadelphia: B.C. Decker (1997).

5.  Llewellyn-Smith, I. J., & Verberne, A. J. M. Central regulation of autonomic functions. New York: Oxford University Press (2011).

6.  Uylings, H. B. M. Cognition, emotion and autonomic responses: The integrative role of the prefrontal cortex and limbic structures. Amsterdam: Elsevier (2000).

7.  Sterling, P. What is health? Allostasis and the evolution of human design. Cambridge, Massachusetts: The MIT Press (2020).

8.  Nieuwenhuys, R. The insular cortex: A review. Progress in Brain Research, 195, 123-164 (2012).

9.  Barrett, L. F. & Simmons, W. K. Interoceptive predictions in the brain. Nature Reviews Neuroscience, 16, 419-429 (2015).

10. Critchley, H. D. & Harrison, N. A. Visceral Influences on Brain and Behavior. Neuron, 77, 624-638 (2013).

11. Thayer, J. F. & Sternberg, E. M. Neural aspects of immunomodulation: Focus on the vagus nerve. Brain, Behavior, and Immunity, 24, 1223-1228 (2010).

12. Craig, A.D. How do you feel? Interoception: the sense of the physiological condition of the body. Nat. Rev. Neurosci. 3, 655–666 (2002).

13. Craig, A. D. How do you feel — now? The anterior insula and human awareness. Nature Reviews Neuroscience, 10, 59-70 (2009).

14. Seth, A. K. & Friston, K. J. Active interoceptive inference and the emotional brain. Philosophical Transactions: Biological Sciences, 371, 1708, 1-10 (2016).

15. McEwen, B. S. Physiology and neurobiology of stress and adaptation: central role of the brain. *Physiological reviews*, *87*, 873-904 (2007).

16. Banks, W. A., & Erickson, M. A. The blood–brain barrier and immune function and dysfunction. *Neurobiology of disease*, *37*, 26-32 (2010).

17. Gibbins, I. L. Peripheral autonomic nervous system. In *The human nervous system* (pp. 93-123). Academic Press New York (1990).

18. Mesulam, M. M., & Mufson, E. J. Insula of the old world monkey. I. Architectonics in the insulo-orbito-temporal component of the paralimbic brain. The Journal of Comparative Neurology, 212, 1-22 (1982).

19. Mufson, E. J., & Mesulam, M. M. Insula of the old world monkey. II: Afferent cortical input and comments on the claustrum. The Journal of Comparative Neurology, 212, 23-37 (1982).

20. Mesulam, M.-M., & Mufson, E. J. Insula of the old world monkey. III: Efferent cortical output and comments on function. The Journal of Comparative Neurology, 212, 38-52 (1982).

21. Stephani, C., Fernandez-Baca, V. G., Maciunas, R., Koubeissi, M., & Lüders, H. O. Functional neuroanatomy of the insular lobe. Brain Structure & Function, 216, 137-49 (2011).

22. Penfield, W., & Faulk, M. E. The insula: Further observations on its function. Londres (1955).

23. Pugnaghi, M., Meletti, S., Castana, L., Francione, S., Nobili, L., Mai, R., & Tassi, L. Features of somatosensory manifestations induced by intracranial electrical stimulations of the human insula. Clinical Neurophysiology, 122, 2049-2058 (2011).

24. Narayana, S., Laird, A. R., Tandon, N., Franklin, C., Lancaster, J. L., & Fox, P. T. Electrophysiological and functional connectivity of the human supplementary motor area. *Neuroimage*, *62*, 250-265 (2012).

25. Santiesteban, I., Banissy, M. J., Catmur, C., & Bird, G. Enhancing social ability by stimulating right temporoparietal junction. *Current Biology*, *22*, 2274-2277 (2012).

26. Silani, G., Lamm, C., Ruff, C. C., & Singer, T. Right supramarginal gyrus is crucial to overcome emotional egocentricity bias in social judgments. *Journal of neuroscience*, *33*, 15466-15476 (2013).

27. Carter, R. M., & Huettel, S. A. A nexus model of the temporal–parietal junction. *Trends in cognitive sciences*, *17*, 328-336 (2013).

28. Seltzer, B., & Pandya, D. N. Post-rolandic cortical projections of the superior temporal sulcus in the rhesus monkey. *Journal of Comparative Neurology*, *312*, 625-640 (1991).

29. Mutschler, I., Wieckhorst, B., Kowalevski, S., Derix, J., Wentlandt, J., Schulze-Bonhage, A., & Ball, T. Functional organization of the human anterior insular cortex. *Neuroscience letters*, *457*, 66-70 (2009).

30. Nomi, J. S., Schettini, E., Broce, I., Dick, A. S., & Uddin, L. Q. Structural connections of functionally defined human insular subdivisions. *Cerebral Cortex*, *28*, 3445-3456 (2018).

31. Kurth, F., Zilles, K., Fox, P. T., Laird, A. R., & Eickhoff, S. B. A link between the systems: functional differentiation and integration within the human insula revealed by meta-analysis. *Brain Structure and Function*, *214*, 519-534 (2010).

32. Kurth, F., Eickhoff, S. B., Schleicher, A., Hoemke, L., Zilles, K., & Amunts, K. Cytoarchitecture and probabilistic maps of the human posterior insular cortex. *Cerebral Cortex*, *20*, 1448-1461 (2010).

33. Cerliani, L. et al. Probabilistic tractography recovers a rostrocaudal trajectory of connectivity variability in the human insular cortex. *Human brain mapping*, *33*, 2005-2034 (2012).

34. Tanaka, S. C., Doya, K., Okada, G., Ueda, K., Okamoto, Y., & Yamawaki, S. Prediction of immediate and future rewards differentially recruits cortico-basal ganglia loops. *Nature neuroscience*, *7*, 887-893 (2004).

35. Chikama, M., McFarland, N. R., Amaral, D. G., & Haber, S. N. Insular cortical projections to functional regions of the striatum correlate with cortical cytoarchitectonic organization in the primate. *Journal of Neuroscience*, *17*, 9686-9705 (1997).

36. Haber, S. N., Fudge, J. L., & McFarland, N. R. Striatonigrostriatal pathways in primates form an ascending spiral from the shell to the dorsolateral striatum. *Journal of Neuroscience*, *20*, 2369-2382 (2000).

37. Alexander, G. E., DeLong, M. R., & Strick, P. L. Parallel organization of functionally segregated circuits linking basal ganglia and cortex. *Annual review of neuroscience*, *9*, 357-381 (1986).

38. Calabresi, P., Picconi, B., Tozzi, A., Ghiglieri, V., & Di Filippo, M. Direct and indirect pathways of basal ganglia: a critical reappraisal. *Nature neuroscience*, *17*, 1022 (2014).

39. Friston, K., Kilner, J., & Harrison, L. A free energy principle for the brain. *Journal of Physiology-Paris*, *100*, 70-87 (2006).

40. Friston, K. Hierarchical models in the brain. *PLoS computational biology*, *4*(11) (2008).

41. Friston, K. The free-energy principle: a unified brain theory?. *Nature reviews neuroscience*, *11*, 127-138 (2010).

42. Friston, K., Schwartenbeck, P., FitzGerald, T., Moutoussis, M., Behrens, T., & Dolan, R. J. The anatomy of choice: active inference and agency. *Frontiers in human neuroscience*, *7*, 598 (2013).

43. Friston, K. J., Daunizeau, J., & Kiebel, S. J. Reinforcement learning or active inference?. *PLoS one*, *4*(7) (2009).

44. Rao, R. P., & Ballard, D. H. Predictive coding in the visual cortex: a functional interpretation of some extra-classical receptive-field effects. *Nature neuroscience*, *2*, 79-87 (1999).

45. Rao, R. P., & Sejnowski, T. J. Predictive Coding, Cortical Feedback, and Spike-Timing Dependent Plasticity. *Probabilistic models of the brain*, 297 (2002).

46. Hikosaka, O. et al. Parallel neural networks for learning sequential procedures. *Trends in neurosciences*, *22*, 464-471 (1999).

47. Hikosaka, O., Ghazizadeh, A., Griggs, W., & Amita, H. Parallel basal ganglia circuits for decision making. *Journal of Neural Transmission*, *125*, 515-529 (2018).

48. Doya, K. What are the computations of the cerebellum, the basal ganglia and the cerebral cortex?. *Neural networks*, *12*, 961-974 (1999).

49. Belin, D., Jonkman, S., Dickinson, A., Robbins, T. W., & Everitt, B. J. Parallel and interactive learning processes within the basal ganglia: relevance for the understanding of addiction. *Behavioural brain research*, *199*, 89-102 (2009).

50. Jueptner, M., Stephan, K. M., Frith, C. D., Brooks, D. J., Frackowiak, R. S., & Passingham, R. E. Anatomy of motor learning. I. Frontal cortex and attention to action. *Journal of neurophysiology*, *77*, 1313-1324 (1997).

51. Jueptner, M., Frith, C. D., Brooks, D. J., Frackowiak, R. S. J., & Passingham, R. E. Anatomy of motor learning. II. Subcortical structures and learning by trial and error. *Journal of neurophysiology*, *77*, 1325-1337 (1997).

52. Valentin, V. V., Dickinson, A., & O'Doherty, J. P. Determining the neural substrates of goal-directed learning in the human brain. *Journal of Neuroscience*, *27*, 4019-4026 (2007).

53. Fermin, A. S., Yoshida, T., Yoshimoto, J., Ito, M., Tanaka, S. C., & Doya, K. Model-based action planning involves cortico-cerebellar and basal ganglia networks. *Scientific reports*, *6*, 31378 (2016).

54. Daw, N. D., O'doherty, J. P., Dayan, P., Seymour, B., & Dolan, R. J. Cortical substrates for exploratory decisions in humans. *Nature*, *441*, 876-879 (2006).

55. Daw, N. D., Niv, Y., & Dayan, P. Uncertainty-based competition between prefrontal and dorsolateral striatal systems for behavioral control. *Nature neuroscience*, *8*, 1704-1711 (2005).

56. Gläscher, J., Daw, N., Dayan, P., & O'Doherty, J. P. States versus rewards: dissociable neural prediction error signals underlying model-based and model-free reinforcement learning. *Neuron*, *66*, 585-595 (2010).

57. Wunderlich, K., Dayan, P., & Dolan, R. J. Mapping value based planning and extensively trained choice in the human brain. *Nature neuroscience*, *15*, 786 (2012).

58. Hasselmo, M. E., & Schnell, E. Laminar selectivity of the cholinergic suppression of synaptic transmission in rat hippocampal region CA1: computational modeling and brain slice physiology. *Journal of Neuroscience*, *14*, 3898-3914 (1994).

59. Hasselmo, M. E. The role of acetylcholine in learning and memory. *Current opinion in neurobiology*, *16*, 710-715 (2006).

60. Hasselmo, M. E., & Sarter, M. Modes and models of forebrain cholinergic neuromodulation of cognition. *Neuropsychopharmacology*, *36*, 52-73 (2011).

61. Chouchou, F. et al. How the insula speaks to the heart: Cardiac responses to insular stimulation in humans. *Human Brain Mapping*, *40*, 2611-2622 (2019).

62. Oppenheimer, S. M., Wilson, J. X., Guiraudon, C., & Cechetto, D. F. Insular cortex stimulation produces lethal cardiac arrhythmias: a mechanism of sudden death? *Brain research*, *550*, 115-121 (1991).

63. Watabe-Uchida, M., Zhu, L., Ogawa, S. K., Vamanrao, A., & Uchida, N. Whole-brain mapping of direct inputs to midbrain dopamine neurons. *Neuron*, *74*, 858-873 (2012).

64. Matsumoto, M., & Hikosaka, O. Two types of dopamine neuron distinctly convey positive and negative motivational signals. *Nature*, *459*, 837-841 (2009).

65. Bromberg-Martin, E. S., Matsumoto, M., & Hikosaka, O. Dopamine in motivational control: rewarding, aversive, and alerting. *Neuron*, *68*, 815-834 (2010).

66. Schultz, W., Dayan, P., & Montague, P. R. A neural substrate of prediction and reward. *Science*, *275*, 1593-1599 (1997).

67. Hikosaka, O., Takikawa, Y., & Kawagoe, R. Role of the basal ganglia in the control of purposive saccadic eye movements. *Physiological reviews*, *80*, 953-978 (2000).

68. Alexander, G. E., & Crutcher, M. D. Functional architecture of basal ganglia circuits: neural substrates of parallel processing. *Trends in neurosciences*, *13*, 266-271 (1990).

69. Hikosaka, O. et al. Direct and indirect pathways for choosing objects and actions. *European Journal of Neuroscience*, *49*, 637-645 (2019).

70. Soares-Cunha, C., Coimbra, B., Sousa, N., & Rodrigues, A. J. Reappraising striatal D1-and D2-neurons in reward and aversion. *Neuroscience & Biobehavioral Reviews*, *68*, 370-386 (2016).

71. Jin, X., Tecuapetla, F., & Costa, R. M. Basal ganglia subcircuits distinctively encode the parsing and concatenation of action sequences. *Nature neuroscience*, *17*, 423-430 (2014).

72. Tecuapetla, F., Jin, X., Lima, S. Q., & Costa, R. M. Complementary contributions of striatal projection pathways to action initiation and execution. *Cell*, *166*, 703-715 (2016).

73. Johansen, J. P., Cain, C. K., Ostroff, L. E., & LeDoux, J. E. Molecular Mechanisms of Fear Learning and Memory. *Cell, 147*, 509-524 (2011).

74. Sternson, S. M. Hypothalamic Survival Circuits: Blueprints for Purposive Behaviors. *Neuron, 77*, 810-824 (2013).

75. LeDoux, J. Rethinking the Emotional Brain. *Neuron, 73*, 1052-1052 (2012).

76. Gerozissis, K. Brain insulin, energy and glucose homeostasis; genes, environment and metabolic pathologies. *European journal of pharmacology*, *585*, 38-49 (2008).

77. Johnston, E., & Olson, L. *The feeling brain: The biology and psychology of emotions*. WW Norton & Company (2015).

78. Damasio, A. R. et al. Subcortical and cortical brain activity during the feeling of self-generated emotions. *Nature neuroscience*, *3*, 1049-1056 (2000).

79. Damasio, A., & Carvalho, G. B. The nature of feelings: evolutionary and neurobiological origins. *Nature reviews neuroscience*, *14*, 143-152 (2013).

80. Kowalski, S. D., & Bondmass, M. D. Physiological and psychological symptoms of grief in widows. *Research in nursing & health*, *31*, 23-30 (2008).

81. Kanel, D., Al-Wasity, S., Stefanov, K., & Pollick, F. E. Empathy to emotional voices and the use of real-time fMRI to enhance activation of the anterior insula. *NeuroImage*, *198*, 53-62 (2019).

82. Lucas, M. V., Anderson, L. C., Bolling, D. Z., Pelphrey, K. A., & Kaiser, M. D. Dissociating the neural correlates of experiencing and imagining affective touch. *Cerebral Cortex*, *25*, 2623-2630 (2015).

83. Björnsdotter, M., & Olausson, H. Vicarious responses to social touch in posterior insular cortex are tuned to pleasant caressing speeds. *Journal of Neuroscience*, *31*, 9554-9562 (2011).

84. Haker, H., Kawohl, W., Herwig, U., & Rössler, W. Mirror neuron activity during contagious yawning - an fMRI study. *Brain imaging and behavior*, *7*, 28-34. (2013).

85. Koelsch, S. Brain correlates of music-evoked emotions. *Nature Reviews Neuroscience*, *15*, 170-180 (2014).

86. Brown, S., Gao, X., Tisdelle, L., Eickhoff, S. B., & Liotti, M. Naturalizing aesthetics: brain areas for aesthetic appraisal across sensory modalities. *Neuroimage*, *58*, 250-258 (2011).

87. Smith, A. P., Henson, R. N., Rugg, M. D., & Dolan, R. J. Modulation of retrieval processing reflects accuracy of emotional source memory. *Learning & Memory*, *12*, 472-479 (2005).

88. Fossati, P. et al. Distributed self in episodic memory: neural correlates of successful retrieval of self-encoded positive and negative personality traits. *Neuroimage*, *22*, 1596-1604 (2004).

89. Koechlin, E., Ody, C., & Kouneiher, F. The architecture of cognitive control in the human prefrontal cortex. *Science*, *302*, 1181-1185 (2003).

90. Ito, M., & Doya, K. Multiple representations and algorithms for reinforcement learning in the cortico-basal ganglia circuit. *Current opinion in neurobiology*, *21*, 368-373 (2011).

91. Fleming, S. M., Weil, R. S., Nagy, Z., Dolan, R. J., & Rees, G. Relating introspective accuracy to individual differences in brain structure. *Science*, *329*, 1541-1543 (2010).

92. Smittenaar, P., Prichard, G., FitzGerald, T. H., Diedrichsen, J., & Dolan, R. J. Transcranial direct current stimulation of right dorsolateral prefrontal cortex does not affect model-based or model-free reinforcement learning in humans. *PLoS One*, *9*(1) (2014).

93. Wong, S. W., Massé, N., Kimmerly, D. S., Menon, R. S., & Shoemaker, J. K. Ventral medial prefrontal cortex and cardiovagal control in conscious humans. *Neuroimage*, *35*, 698-708 (2007).

94. Erdogan, E. T., Saydam, S. S., Kurt, A., & Karamursel, S. Anodal transcranial direct current stimulation of the motor cortex in healthy volunteers. *Neurophysiology*, *50*, 124-130 (2018).

95. Pan, X., Sawa, K., Tsuda, I., Tsukada, M., & Sakagami, M. Reward prediction based on stimulus categorization in primate lateral prefrontal cortex. *Nature neuroscience*, *11*, 703 (2008).

96. Rees, G., Frackowiak, R., & Frith, C. Two modulatory effects of attention that mediate object categorization in human cortex. *Science*, *275*, 835-838 (1997).

97. Dolan, R. J., & Fletcher, P. C. Dissociating prefrontal and hippocampal function in episodic memory encoding. *Nature*, *388*, 582-585 (1997).

98. Vikbladh, O. M. et al. Hippocampal contributions to model-based planning and spatial memory. *Neuron*, *102*, 683-693 (2019).

99. Akkal, D., Dum, R. P., & Strick, P. L. Supplementary motor area and presupplementary motor area: targets of basal ganglia and cerebellar output. *Journal of Neuroscience*, *27*, 10659-10673 (2007).

100. Coffman, K. A., Dum, R. P., & Strick, P. L. Cerebellar vermis is a target of projections from the motor areas in the cerebral cortex. *Proceedings of the National Academy of Sciences*, *108*, 16068-16073 (2011).